\documentclass[conference]{IEEEtran}
\IEEEoverridecommandlockouts
\usepackage{cite}
\usepackage{amsmath,amssymb,amsfonts}
\usepackage{algorithmic}
\usepackage{graphicx}
\usepackage{textcomp}
\usepackage{xcolor}

\usepackage{algorithm}
\usepackage{algorithmic}
 
\makeatletter
\newcommand{\removelatexerror}{\let\@latex@error\@gobble}
\makeatother

\def\BibTeX{{\rm B\kern-.05em{\sc i\kern-.025em b}\kern-.08em
    T\kern-.1667em\lower.7ex\hbox{E}\kern-.125emX}}
\begin{document}

\title{Channel Correlation Matrix Extrapolation Based on Roughness Calibration of Scatterers\\
\thanks{The research presented in this paper has been kindly supported by the projects as follows, National Natural Science Foundation of China under Grants (62394294, 62394290), the Fundamental Research Funds for the Central Universities under Grant 2242022k60006. }
}

\author{
    \IEEEauthorblockN{Heling Zhang\IEEEauthorrefmark{1}\IEEEauthorrefmark{3} Xiujun Zhang\IEEEauthorrefmark{2} Xiaofeng Zhong\IEEEauthorrefmark{1}\IEEEauthorrefmark{3} Shidong Zhou\IEEEauthorrefmark{1}\IEEEauthorrefmark{3}}
    \IEEEauthorblockA{\IEEEauthorrefmark{1}Department of Electronic Engineering, Tsinghua University, Beijing, China}
    \IEEEauthorblockA{\IEEEauthorrefmark{2}Beijing National Research Center for Information Science and Technology}
    \IEEEauthorblockA{\IEEEauthorrefmark{3}State Key Laboratory of Space Network and Communications\\
    Emails: zhanghl24@mails.tsinghua.edu.cn, \{zhangxiujun, zhongxf, zhousd\}@tsinghua.edu.cn}
}

\maketitle

\begin{abstract}
To estimate the channel correlation matrix (CCM) in areas where channel information cannot be collected in advance, this paper proposes a way to spatially extrapolate CCM based on the calibration of the surface roughness parameters of scatterers in the propagation scene. We calibrate the roughness parameters of scene scatters based on CCM data in some specific areas. From these calibrated roughness parameters, we are able to generate a good prediction of the CCM for any other area in the scene by performing ray tracing. Simulation results show that the channel extrapolation method proposed in this paper can effectively realize the extrapolation of the CCM between different areas in frequency domain, or even from one domain to another.
\end{abstract}


\section{Introduction}
Current wireless communication systems estimate CSI by sending pilots, which results in considerable resource consumption \cite{1}. An accurate estimation of the channel correlation matrix (CCM) can bring many benefits to communication tasks, such as reducing the pilot overhead and increasing the accuracy of channel estimation. Although the CCM can be estimated by sending pilots to collect channel information, this will lead to huge pilot consumption in the areas where channel information cannot be collected in advance. 

Channel extrapolation means that, using channel state information (CSI), in spatial or frequency domain, within a certain range in a propagation scene to infer CSI within another range \cite{2}. Considering the significant role of CCM in channel estimation, channel extrapolation of CCM is particularly meaningful. However, since CCM reflects the complex interaction between electromagnetic waves and environment scatterers, even in the same propagation scene, the CCM of different areas cannot be extrapolated simply, which increases the difficulty of the channel extrapolation problem.

Previous studies have shown that the roughness of the scatterer surface in the propagation scene largely shapes the characteristics of the channel. \cite{3} shows that when the scatterer surface is rough enough, the electromagnetic waves scattered will form a cluster of multipaths, greatly increasing the number of multipaths and improving the degree of freedom of the MIMO system; \cite{4} shows that the roughness of the scatterers in the propagation scene will affect the channel correlation in different areas; \cite{5} researches on the relationship between the roughness of the scatterer surface and the channel characteristics through simulation, pointing out that the scatterer roughness has an impact on the angluar spread, root mean square delay, and coherence bandwidth of the received signal. Inspired by these studies, by calibrating the roughness of the scatterers in the environment, we can establish a digital twin of real electromagnetic propagation scene, which can provide assistance on CCM extrapolation.

There have been some studies that attempt to calibrate the roughness of environmental scatterers using existing channel information, or to calculate channel parameters by estimating the roughness of scatterers. \cite{6} proposed a parameterized model to describe the roughness of scatterers, and used the estimated roughness of scatterers to calculate the path loss; based on \cite{6}, \cite{7} and \cite{8} calibrated the roughness parameters of common materials. In \cite{9} , the electromagnetic parameters and roughness parameter distribution on the scatterer surface are learned from indoor data based on ray tracing simulation, and the channel gain and delay spread are predicted from the learned distribution. 

However, there is still no work dedicated to extrapolating CCM by calibrating the roughness parameters, which is a statistic that greatly helps channel estimation. To fill this gap, this paper proposes a method to spatially extrapolate CCM, that is, to calibrate the roughness parameters of scatterers using the known channel correlation matrix, and to calculate the channel correlation matrix at other locations based on these calibrated parameters.

The remainder of the paper is organized as follows. Section \ref{Problem Formulation} describes the channel extrapolation problem. Section \ref{Scatterer modeling} parametrically models the surface roughness of the scatterer. Section \ref{Extrapolation algorithm} presents the channel extrapolation algorithm, and Section \ref{Simulation Result} verifies its performance by simulation. The conclusion of the paper is given in Section \ref{Conclusion}.

\section{Problem Formulation}\label{Problem Formulation}
\begin{figure}[htbp]
\centerline{\includegraphics[scale=0.5]{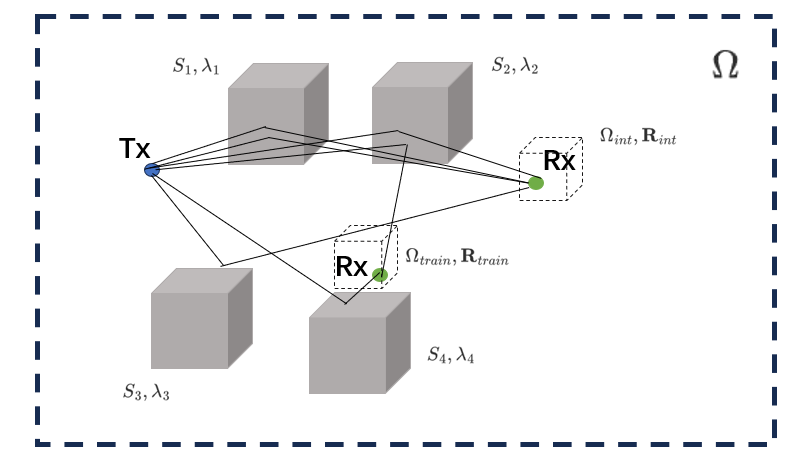}}
\caption{Scene of channel extrapolation}
\label{scene of channel extrapolation}
\end{figure}

As shown in Fig.\ref{scene of channel extrapolation}, in a propagation scene $\Omega$ without a line-of-sight path from the transmitter Tx and the receiver Rx, which contains n scatterers $\{S_1, S_2,...,S_n\}$, the CCM $\mathbf{R}_{int} = f(\Omega_{int};\Lambda)$ of any interested area $\Omega_{int} \subset \Omega$ is mostly decided by the location of $\Omega_{int}$ and the roughness parameters set $\Lambda=\{\mathbf{\lambda_1},\mathbf{\lambda_2}, ...,\mathbf{\lambda_n}\}$ of all n scatterers. Given the CCM $\mathbf{R}_{train}$ of another area $\Omega_{train}$, which does not intersect with the interested area, channel extrapolation is to estimate $\Lambda$ through $\mathbf{R}_{train}$, from which we can derive $\mathbf{R}_{int}$ of any interested area $\Omega_{int}$.

\section{Scatterer modeling}\label{Scatterer modeling}
In our extrapolation algorithm, each scatter $S_i$ is assigned to a roughness parameter vector $\mathbf{\lambda_i}$. This section will present how we model scatterers in the scene and how $\mathbf{\lambda_i}$ is derived from the model.

The surface of a scatterer can be complex, and is usually composed of several kinds of materials with completely different roughness. Taking this into consideration, we divide the surface $\Sigma_i$ of the scatterer $S_i$ into $m_{tile}$ rows and $n_{tile}$ columns, $m_{tile} \times n_{tile}$ tiles in total, each tile has a corresponding positive roughness parameter $\alpha_r$, which will be utilized later in the calculation of scattering channel\cite{10}. Arrange these roughness parameters to obtain a roughness matrix $\mathbf{A} \in \mathbb{R}^{m_{tile} \times n_{tile}}$. 

The roughness parameter $\alpha_r$ should always be positive. To avoid possible difficulty in parameter optimization caused by this constraint, suppose $\alpha_r = g(z), g:\mathbb{R} \rightarrow \mathbb{R^+}$, and $\mathbf{A}$ is mapped element-wisely from $\mathbf{Z}$ through function $g$. Considering that $\alpha_r$ is generally not too large, $g$ is set as sigmoid function with a maximum of 15 in our algorithm. In principle, it can be any bijection between $\mathbb{R}$ and $\mathbb{R^+}$ such as the exponential function.

In order to reduce the number of parameters to be estimated, principal component analysis is performed on matrix $\mathbf{Z}$. Assume that all elements of the matrix $\mathbf{Z}$ follow a joint Gaussian distribution, that is, let $T_1,T_2$ are two tiles generated by dividing the surface, then their corresponding parameters $z_1,z_2$ satisfy
\begin{equation}
    z_1 \sim N(0,1),\ z_2\sim N(0,1)\label{eq1}
\end{equation}

\begin{equation}
E(z_1z_2)=e^{-\frac{d(T_1,T_2)}{l_{corr}}}\label{eq2}
\end{equation}
where $d(T_1,T_2)$ is the distance between the centers of the two tiles; $l_{corr}$ is the correlation length, indicating the severity of the change in the roughness (in other words, the frequency of material changes) of the scatterer surface. Flatten the matrix $Z$ into $vec(\mathbf{Z})$, then $vec(\mathbf{Z})$ is a Gaussian random vector, and the corresponding correlation matrix is
\begin{equation}
    \mathbf{R_z} = E(vec(\mathbf{Z})vec(\mathbf{Z})^T)\label{eq3}
\end{equation}
$\mathbf{R_z}=\mathbf{Q}\mathbf{\Lambda_{eig}} \mathbf{Q}^T$. Keep the $k_{eig}$ largest eigen values and set others to 0, $vec{(\mathbf{Z}})$ can be approximated by a linear combination of $k_{eig}$ independent principal components. For scatterer $S_i$, the corresponding roughness vector $vec{(\mathbf{Z_i})}$ can be approximated as
\begin{equation}
    vec(\mathbf{Z_i}) \approx \tilde{\mathbf{Q_i}}\tilde{\mathbf{\Lambda}}_{eig,i}^{\frac{1}{2}}\mathbf{\lambda_i}\label{eq4}
\end{equation}
where $\tilde{\mathbf{\Lambda}}_{eig,i}\in \mathbb{R}^{k_{eig} \times k_{eig}}$ contains the largest $k_{eig}$ eigen values in $\mathbf{\Lambda}_{eig}$, and $\tilde{\mathbf{Q_i}} \in \mathbb{R}^{k_{eig}\times(m_{tile}\times n_{tile})}$ is the corresponding vectors. In this way, the roughness vector $\lambda_i$ corresponding to the scatterer $S_i$ is derived, and all n roughness vectors $\{\mathbf{\lambda_1},\mathbf{\lambda_2}, ...,\mathbf{\lambda_n}\}$ constitute the parameter set $\Lambda$ to be estimated.

\section{Extrapolation algorithm}\label{Extrapolation algorithm}
We obtain the optimal value of the parameter $\Lambda$ by executing gradient descent. This section will show how to obtain the estimated value of the CCM $\hat{\mathbf{R}}_{train}$ from the parameter $\Lambda$, and how to calculate the loss between the estimated value and the ground truth.
\subsection{Approximating CCM by Ray Tracing}
In the propagation scene $\Omega$ with a fixed transmitter Tx at $\mathbf{r}_{Tx}$, for receiver Rx at any possible position $\mathbf{r}_{Rx}\in\Omega$, the channel parameters (frequency response or antenna response matrix in MIMO system) of the link between them can be approximated by ray tracing simulation. Ray tracing approximates the electromagnetic waves radiated by Tx with a large number of rays emitted from Tx. After being scattered by scatterers in the propagation scene, a total of $n_{ray}$ rays will eventually reach Rx. In frequency domain, the approximate channel response generated by ray tracing is
\begin{equation}
    H(f) = \sum_{i=0}^{n_{ray}}a_ie^{-j2\pi f\tau_i+\phi}\label{eq5}
\end{equation}
for any sub-carrier frequency $f$, where $a_i$ and $\tau_i$ are the complex gain and delay of the ith path respectively. $\phi$ is a random phase uniformly distributed between $[0,\pi]$. 

While $\tau_i$ is simply determined by the path length corresponding to the ray, the factors affecting complex path gain $a_i$ are more complicated. Besides path loss, $a_i$ is also affected by the scattering attenuation on the surface of each scatterer. For an incident ray on the surface of a scatterer, suppose that the intersection of the ray and the surface is $P$, the unit vector on scattering direction is $\hat{\mathbf{k_s}}$, the unit vector on the reflected ray direction is $\hat{\mathbf{k_r}}$, the reflected electric field calculated by the reflection law is $E_r$, and the point $P$ is located on a small block with a roughness parameter of $\alpha_r$, then the electric field corresponding to the scattered ray is calculated by Gaussian scattering model \cite{10}
\begin{equation}
    E_s = \frac{1}{a(\alpha_r)}E_re^{\alpha_r(\hat{\mathbf{k_r}} \cdot \hat{\mathbf{k_s}}-1)}\label{eq6}
\end{equation}
where $\alpha_r \ge 0$ describes the roughness degree of the surface around $P$, $a(\alpha_r)$ is a normalization factor. As the plane becomes smoother, $\alpha_r$ becomes larger, and the energy of the scattered electromagnetic waves becomes more concentrated. Note that $\mathbf{A_i}$ derived from $\lambda_i$ describes all the roughness information of scatterer $S_i$. Through ray tracing simulation, the channel response $H(f)$ at the position $\mathbf{r}_{Rx}$ depends on the roughness distribution $\mathbf{A}$ of the scatterer surface through the complex gain $a$ of each ray, and eventually depends on the parameter set $\Lambda$.

Furthermore, the CCM in the area $\Omega_{train}$ can be approximately calculated by Monte Carlo method. We randomly select $N_{MC}$ points in $\Omega_{train}$, and calculate the channel responses of the receiver at these positions $\hat{\bf{H}}_{train}=[\hat{\bf{h}}_1, \hat{\bf{h}}_2, ..., \hat{\bf{h}}_{N_{MC}}]$ from current estimated values of the parameters $\hat{\Lambda}$. The correlation matrix can be approximated as $\hat{\bf{R}}_{train}=\frac{1}{N_{MC}}{\hat{\bf{H}}_{train}\hat{\bf{H}}^H_{train}}$. Though the $\hat{\bf{R}}_{train}$ discussed above is in frequency domain, for the correlation matrix of antenna response in spatial domain, the principle is similar.

As an additional explanation, we should note that the complex gain $a$ of a path is also influenced by the complex permittivity $\eta$ of the scene scatterers. However, compared with the roughness parameter, the complex permittivity has a much less impact on the multipath gain, making estimating them not only difficult but also unnecessary. Therefore, when calculating the channel response and correlation matrix, we simply preset a typical value $\hat{\eta}$ for complex permittivities of all scatterers in the scene. 

\subsection{Loss Function}
Given that we use gradient descent to optimize the roughness parameter $\Lambda$, we need to define a loss function between the estimation of CCM $\hat{\mathbf{R}}_{train}(\Lambda)$ and the ground truth $\mathbf{R}_{train}$. We do not limit the domain of the correlation matrix used for training: $\mathbf{R}_{train}$ can reflect the correlation of frequency domain responses, in which case $\mathbf{R}_{train}\in\mathbb{C}^{n_{freq}\times n_{freq}}$; it can also reflect the correlation between transmitter or receiver antenna responses, in which case $\mathbf{R}_{train} \in \mathbb{C}^{n_{Tx}\times n_{Tx}}$ or $\mathbf{R}_{train} \in \mathbb{C}^{n_{Rx}\times n_{Rx}}$, where $n_{freq}$, $n_{Tx}$, and $n_{Rx}$ are the number of subcarriers used for training, the number of transmitter and receiver antennas respectively.

For the case where $\mathbf{R}_{train}$ is in the frequency domain, we define the loss function to minimize the channel estimation error under MMSE algorithm. Consider a SISO-OFDM system, the channel model is
\begin{equation}
    \mathbf{y} = \mathbf{X}\mathbf{h}+\mathbf{n}\label{eq7}
\end{equation}

where $\mathbf{h}$ is the frequency domain response, $\mathbf{X}$ is the pilot matrix, and $\mathbf{n}$ is random noise with an average energy of $\sigma_n^2$. The estimated value of the frequency domain response $\hat{\mathbf{h}}$ is
\begin{equation}
    \hat{\mathbf{h}} = \mathbf{W}\mathbf{y}\label{eq8}
\end{equation}
where $ \mathbf{W}=\mathbf{R}_{train}(\mathbf{X}^H\mathbf{X}\mathbf{R}_{train}+\sigma_n^2\mathbf{I})^{-1}\mathbf{X}^H$ is the filter matrix. In our case, when CCM $\mathbf{R}_{train}$ has an estimation error, $\mathbf{R}_{train}$ in the MMSE filter matrix $\mathbf{W}$ is replaced by its estimation $\hat{\mathbf{R}}_{train}$. The mean square error of the frequency domain response estimation is
\begin{equation}
    \epsilon = E(|\hat{\mathbf{W}}\mathbf{y}-\mathbf{h}|^2)\label{eq9}
\end{equation}
where $\hat{\mathbf{W}}=\hat{\mathbf{R}}_{train}(\mathbf{X}^H\mathbf{X}\hat{\mathbf{R}}_{train}+\sigma_n^2\mathbf{I})^{-1}\mathbf{X}^H$. For the sake of simplicity, let pilot matrix $\mathbf{X}$ be an identity matrix, the mean square error of channel estimation can be simplified as
\begin{equation}
    \epsilon = \frac{1}{SNR}tr(\mathbf{R}_{train}\mathbf{R}_y^{-1})+\frac{1}{SNR^2}tr(\mathbf{R}_{train}(\mathbf{R}_y^{-1}-\hat{\mathbf{R}}_y^{-1})^2)\label{eq10}
\end{equation}
where $SNR=\frac{1}{\sigma_n^2}$, $\mathbf{R}_y=\mathbf{R}_{train}+\frac{1}{SNR}\mathbf{I}$, $\hat{\mathbf{R}}_y=\hat{\mathbf{R}}_{train}+\frac{1}{SNR}\mathbf{I}$. The first term of \eqref{eq9} depends on the propagation scene itself, while the second term depends on the estimation accuracy of the channel correlation matrix. To minimize the channel estimation error, we set loss function to be
\begin{equation}
    \mathcal{L}_1(\mathbf{R}_{train}, \hat{\mathbf{R}}_{train})  = \frac{1}{SNR^2}tr(\mathbf{R}_{train}(\mathbf{R}_y^{-1}-\hat{\mathbf{R}}_y^{-1})^2)\label{eq11}
\end{equation}

For the case of spatial domain, since the closed form of CCM estimation error under MMSE algorithm is not easy to derive, we simply use the mean square error between the ground truth and the estimation as the error function:
\begin{equation}
    \mathcal{L}_2=||\mathbf{R}_{train}-\hat{\mathbf{R}}_{train}||_F^2\label{eq12}
\end{equation}

\subsection{Extrapolation Algorithm}
After illustrating the way to calculate CCM and defining the loss function, we can finally describe our extrapolation algorithm in Algorithm 1.

\begin{figure}[!t]
    \label{alg:1}
    \removelatexerror
    \begin{algorithm}[H]
        \caption{CCM Extrapolation Algorithm}
        \begin{algorithmic}[1]
            \REQUIRE Training area $\Omega_{train}$, channel correlation matrix $\mathbf{R}_{train}$ in the training area, interested area $\Omega_{int}$        
            \ENSURE Estimated CCM $\hat{\mathbf{R}}_{int}$ in the interested area 
            \STATE set $l_{corr}$, function $g(z)$, $\hat{\eta}$, $m_{tile}$ and $n_{tile}$ for each scatter in the scene.
            \STATE set Monte Carlo parameter $N_{MC}$.
            \STATE randomly initialize parameter set $\Lambda$.
            \REPEAT 
            \STATE Calculate $\hat{\mathbf{R}}_{train}=f(\Omega_{train};\Lambda)$ in $\Omega_{train}$ by ray tracing.
            \STATE $Loss(\Lambda)=\mathcal{L}(\hat{\mathbf{R}}_{train},\mathbf{R}_{train})$
            \STATE $d\Lambda$ = $\nabla_{\Lambda}Loss(\Lambda)$
            \STATE $\Lambda \leftarrow \Lambda + d\Lambda$
            \UNTIL{convergence}
            \STATE Calculate $\hat{\mathbf{R}}_{int}=f(\Omega_{int};\Lambda)$ in $\Omega_{int}$ by ray tracing.
        \end{algorithmic}
    \end{algorithm}
\end{figure}

\section{Simulation Result}\label{Simulation Result}
\begin{figure}[htbp]
\centerline{\includegraphics[scale=0.35]{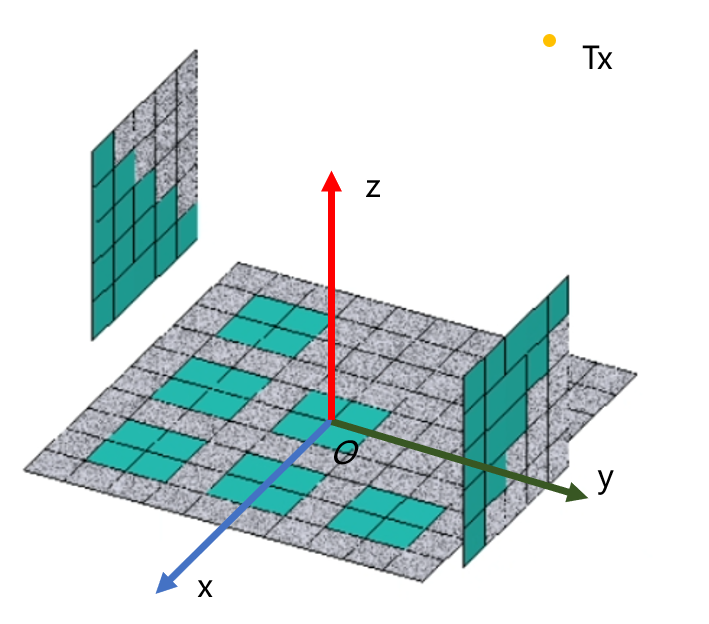}}
\caption{The propagation scene}
\label{Scene Schemetic}
\end{figure}

The feasibility of the proposed channel extrapolation method is evaluated in a simple propagation scene as Fig.\ref{Scene Schemetic}. The code for simulation is publicly available at https://github.com/zhanghl24/CCM-Extrapolation.

In the propagation scene, the transmitter is located at (-50, 0, 50) (unit:m, the same below), which has a single half-wave dipole antenna placed along the x=z direction; the carrier frequency $f_c$ is 3GHz, the communication bandwidth $BW$ is 150MHz. The voltage across the transmitting antenna is 1V.

There are three scatterers in the scene:
\begin{itemize}
    \item $S_1$ is located in the $xOy$ plane, centered at the origin, and consists of 10*10 square blocks with a side length of 3.06;
    \item $S_2$ is parallel to the $xOz$ plane, centered at (7, 18, 9), and consists of 5*5 square blocks with a side length of 5;
    \item $S_3$ is parallel to the $xOz$ plane, centered at (-7, -18, 9), and consists of 5*5 square blocks with a side length of 5.
\end{itemize}

There are two types of square blocks that make up the scatterer:
\begin{itemize}
    \item Concrete: rough surface, $\alpha_r$=1e-3,  $\eta$=6.31-0.26j.
    \item Glass: smooth surface, $\alpha_r$=10, $\eta$=5.24-0.34j.
\end{itemize}
Each scatterer surface is designed to be asymmetric, so that different areas within the propagation scene face very different propagation environments, as is often the case in reality.

We arbitrarily choose $\Omega_{train}$ to be a cubic area centered at (9.38, 7.32, 2.40) with a side length of 3, and $\Omega_{test}$ to be a cubic area centered at (3.76, 11.30, 4.71) with the same side length. By ray tracing simulation, we generate ground truth $\mathbf{R}_{train}$ and $\mathbf{R}_{test}$ in the two area for training and testing. $\hat{\eta}$ is preset to 5.2-0.2j. In addition, we use Adam optimizer\cite{11} to optimize $\Lambda$, and sionna\cite{12} to perform ray tracing. Line-of-sight path is removed from the simulation result as we only consider non-line-of-sight propagation in our algorithm.

\begin{figure}[htbp]
\centerline{\includegraphics[scale=1.1]{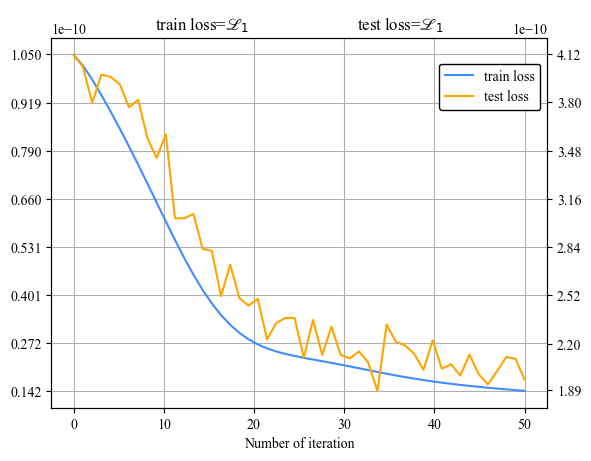}}
\caption{Extrapolation in frequency domain}
\label{f2f}
\end{figure}

Firstly, we use the CCM in frequency domain to perform channel extrapolation. We evenly set 11 subcarriers within $[f-0.5BW, f+0.5BW]$ and calculate the correlation matrix $\mathbf{R}_{train}$ of the channel response on these subcarriers. Using $\mathbf{R}_{train}$ to extrapolate the CCM of the same 11 subcarrier responses in $\Omega_{test}$, the result is shown in Fig.\ref{f2f}.

According to Fig.\ref{f2f}, through channel extrapolation, The variable part of the channel prediction error is reduced by about 50\% in the area $\Omega_{test}$. Although the training area $\Omega_{train}$ and testing area $\Omega_{test}$ do not intersect each other, a good estimate of $\mathbf{R}_{test}$ can still be given by the scatterer parameters learned from $\mathbf{R}_{train}$.

\begin{figure}[htbp]
\centerline{\includegraphics[scale=1.1]{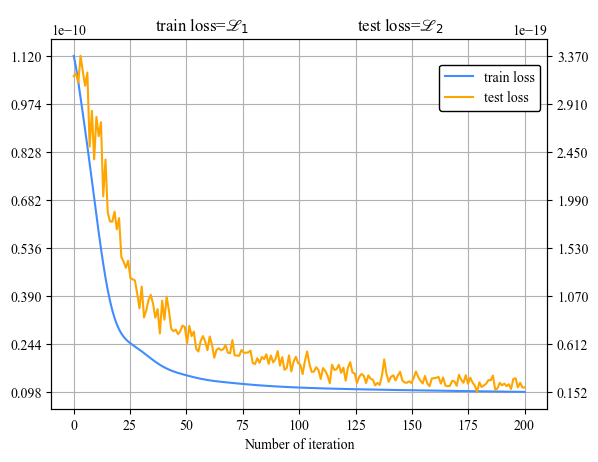}}
\caption{Extrapolation from frequency domain to spatial domain}
\label{f2s}
\end{figure}

We further expand the transmitter and receiver antennas in the scenario to $4 \times 1$ linear array. Still taking the frequency domain correlation matrix $\mathbf{R}_{train}$ as input, we then try to extrapolate the correlation matrix between receiver antenna responses within the region $\Omega_{test}$. The result is shown in Fig.\ref{f2s}. For CCM in frequency domain and spatial domain, the loss is calculated by $\mathcal{L}_1$ in \eqref{eq11} and $\mathcal{L}_2$ in \eqref{eq12}, respectively.

As can be seen from Fig.\ref{f2s}, after iteration, the prediction error of the correlation matrix drops to below 20\% of the original value. Even if the CCM used for training and the target CCM are not in the same domain, a good extrapolation result can still be acheived since our channel extrapolation method is based on the calibration of real scatterer parameters.

\begin{figure}[htbp]
\centerline{\includegraphics[scale=1.1]{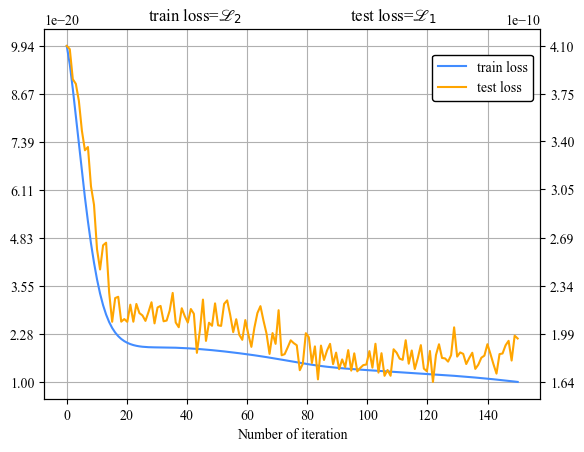}}
\caption{Extrapolation from spatial domain to frequency domain}
\label{s2f}
\end{figure}

We then let $\mathbf{R}_{train}$ be the spatial domain correlation matrix and extrapolate the frequency domain correlation matrix in $\Omega_{test}$, and Fig.\ref{s2f} illustrates the result.

Though it takes longer to converge, extrapolating using spatial CCM reduces the variable channel estimation error in the test area by about 50\%, which is comparable to the result using frequency domain CCM. The longer converge process may be due to the fact that the spatial domain contains less information about the surface roughness of the scatterer in the scene.

\section{Conclusion}\label{Conclusion}
This paper proposes a channel extrapolation method based on calibrating the roughness parameters of the scatterer surface in the environment. We estimate the CCM in a certain area from the roughness parameters through ray tracing, and use the ground truth of the CCM in this area to optimize the roughness parameters. From the optimized parameters, the CCM in the area of interest can be well estimated. Simulation results show that our channel extrapolation method can not only realize the extrapolation of correlation matrix between different regions in the same domain, but also work in different domains. Specifically, the algorithm performs particularly well in the extrapolation from frequency domain CCM to spatial domain CCM. Further research will focus on pilots design based on the estimated CCM to reduce pilot consumption, and to improve channel estimation accuracy.


\begin{thebibliography}{00}
\bibitem{1} F. Fuschini, M. Zoli, E. M. Vitucci, M. Barbiroli, and V. Degli-Esposti, “A study on Millimeter-Wave multiuser directional beamforming based on measurements and ray tracing simulations,” IEEE Transactions on Antennas and Propagation, vol. 67, no. 4, pp. 2633–2644, Apr. 2019, doi: 10.1109/tap.2019.2894271.
\bibitem{2} S. Zhang, Y. Liu, F. Gao, C. Xing, J. An, and O. A. Dobre, “Deep Learning based channel Extrapolation for Large-Scale Antenna Systems: Opportunities, Challenges and Solutions,” IEEE Wireless Communications, vol. 28, no. 6, pp. 160–167, Dec. 2021, doi: 10.1109/mwc.001.2000534.
\bibitem{3} Z. Xu, X. Dong, and J. Bornemann, “A statistical model for the MIMO channel with rough reflection surfaces in the THz band,” Nano Communication Networks, vol. 8, pp. 25–34, Jun. 2016, doi: 10.1016/j.nancom.2015.09.002.
\bibitem{4} N. W. Xu, S. A. Zekavat, and N. H. Tong, “A novel spatially correlated multiuser MIMO channel modeling: impact of surface roughness,” IEEE Transactions on Antennas and Propagation, vol. 57, no. 8, pp. 2429–2438, Aug. 2009, doi: 10.1109/tap.2009.2024491.
\bibitem{5} A. Torabi and S. A. R. Zekavat, “MIMO channel characterization over random rough dielectric terrain,” in *2014 IEEE 25th Annual International Symposium on Personal, Indoor, and Mobile Radio Communication (PIMRC)*, Washington DC, USA: IEEE, Sep. 2014, pp. 161–165. doi: [10.1109/PIMRC.2014.7136152]

\bibitem{6} V. Degli-Esposti, F. Fuschini, E. M. Vitucci, and G. Falciasecca, “Measurement and modelling of scattering from buildings,” IEEE Transactions on Antennas and Propagation, vol. 55, no. 1, pp. 143–153, Jan. 2007, doi: 10.1109/tap.2006.888422.
\bibitem{7} H. Song, Y. Wang, X. Liao and J. Zhang, "Diffuse Scattering Characteristics of Building Materials at Sub-THz Band," 2023 IEEE 11th Asia-Pacific Conference on Antennas and Propagation (APCAP), Guangzhou, China, 2023, pp. 1-2, doi: 10.1109/APCAP59480.2023.10469755. 
\bibitem{8} J. Pascual-Garcia, J.-M. Molina-Garcia-Pardo, M.-T. Martinez-Ingles, J.-V. Rodriguez, and N. Saurin-Serrano, “On the importance of diffuse scattering model parameterization in indoor wireless channels at MM-Wave frequencies,” IEEE Access, vol. 4, pp. 688–701, Jan. 2016, doi: 10.1109/access.2016.2526600.
\bibitem{9} J. Hoydis *et al.*, “Learning Radio Environments by Differentiable Ray Tracing,” Nov. 30, 2023, *arXiv*: arXiv:2311.18558. Accessed: Aug. 20, 2024. [Online]. Available: http://arxiv.org/abs/2311.18558
\bibitem{10} L. J. Chmielewski, R. Kozera, A. Orłowski, K. Wojciechowski, A. M. Bruckstein, and N. Petkov, Eds., *Computer Vision and Graphics: International Conference, ICCVG 2018, Warsaw, Poland, September 17 - 19, 2018, Proceedings*, vol. 11114. in Lecture Notes in Computer Science, vol. 11114. Cham: Springer International Publishing, 2018. doi: [10.1007/978-3-030-00692-1](https://doi.org/10.1007/978-3-030-00692-1).
\bibitem{11} D. P. Kingma and J. Ba, “Adam: A Method for Stochastic Optimization,” Jan. 29, 2017, *arXiv*: arXiv:1412.6980. Accessed: Aug. 20, 2024. [Online]. Available: http://arxiv.org/abs/1412.6980
\bibitem{12} J. Hoydis *et al.*, “Sionna: An Open-Source Library for Next-Generation Physical Layer Research,” Mar. 20, 2023, *arXiv*: arXiv:2203.11854. Accessed: Aug. 20, 2024. [Online]. Available: http://arxiv.org/abs/2203.11854

\end{thebibliography}
\end{document}